\documentclass[aps,pre,preprint,groupedaddress]{revtex4-1}

\usepackage{amsmath}  

\def\firstauthor#1{\gdef\@firstauthor{*}}

\usepackage{color} 
\usepackage{overpic}
\usepackage{amssymb}
\usepackage{hyperref}
\usepackage{inputenc}

\usepackage{graphicx}
\usepackage{caption}
\usepackage[justification=justified,singlelinecheck=false]{caption}

\usepackage{ragged2e}  

\begin{document}


 \title{Optimal dismantling of directed networks}

  \author{Xueming Liu$^{1*}$, Jiawen Hu$^{1}$, Yumei Wang$^{1}$, Yang-Yu Liu$^{2*}$, Hai-Tao Zhang$^{1*}$ }

 \affiliation{
 $^1$ School of Artificial Intelligence and Automation, Engineering Research Center of Autonomous Intelligent Unmanned Systems, and State Key Laboratory of Digital Manufacturing Equipments and Technology, Huazhong University of Science and Technology, Wuhan, China\\
 $^2$ Channing Division of Network Medicine, Brigham and Women’s Hospital and Harvard Medical School, Boston, MA, USA\\ 
 $*$ email: xm\_liu@hust.edu.cn; yyl@channing.harvard.edu; zht@mail.hust.edu.cn\\ 
 }
\vspace{-0.2cm}
\date{\today}

\begin{abstract}

As a fundamental problem in network science, network dismantling focuses on identifying a set of critical nodes whose removal sharply reduces a network's connectivity and functionality. Potential applications include stopping rumor spread, blocking sentiment propagation, and controlling epidemics and pandemics. Previous studies have mainly focused on undirected networks, whereas many real-world networks are inherently directed, such as the World Wide Web and the global trade system. Moreover, the functionality of directed networks depends on the giant strongly connected component (GSCC), where nodes are mutually reachable. Considering both the directionality and heterogeneity of these networks, we propose a novel centrality measure—network incoherence (NI) centrality—and develop a trophic analysis-based dismantling (TAD) method, in which nodes are removed in descending order according to their NI centrality scores, aiming to efficiently dismantle directed networks by reducing the GSCC. When applied to a wide range of benchmark synthetic networks with varying degree heterogeneity and 15 real-world directed networks, our TAD method consistently outperforms existing state-of-the-art methods. Significantly, TAD also induces the largest maximum avalanches during the dismantling process, highlighting its ability to capture structurally critical nodes. These findings provide new insight into the structure-function relationship of directed networks and inform the design of more resilient systems against perturbations.
\end{abstract}

\maketitle

\section{Introduction}

Network dismantling (ND) involves finding the minimum set of critical nodes whose removal can rapidly fragment a network into isolated clusters \cite{CI, fan2020universal}. ND has attracted considerable attention \cite{b13,cohen2000resilience,gnd, kinney2005modeling, achlioptas2009explosive, d2015anomalous} due to its extensive real-world applications, such as stopping the spread of epidemics and rumors \cite{moreno2002epidemic, pastor2001epidemic, moreno2004dynamics}, and disrupting the communication in criminal or malware networks \cite{ribeiro2018dynamical,bpd}. 

To identify critical nodes in networks, various heuristic methods based on centrality measures have been proposed. At the individual node level, nodes are assessed based on centrality measures such as degree \cite{albert2000error, cohen2001breakdown}, closeness \cite{b19}, and betweenness \cite{b20}. At the level of node sets, the Collective Influence (CI) algorithm \cite{CI}, which leverages optimal percolation theory, has been developed to identify key groups of nodes. Additionally, advanced algorithms like Belief Propagation-Guided Decimation (BPD) \cite{bpd} and the Min-Sum algorithm \cite{minsum} address the optimal decycling problem, providing a more refined approach to pinpointing a minimal, yet highly influential set of critical nodes.




Recently, several machine learning-based methods have been developed for network dismantling. Notably, the deep reinforcement learning-based FINDER \cite{finder}, the deep learning-based Directed Network Disintegrator (DND) \cite{zhang2023dnd}, and the supervised learning-based GDM \cite{gdm} have shown potential in handling the complexities of large-scale networks. However, most of these methods are primarily designed for undirected networks. Among them, DND is the only approach tailored for directed networks, but its performance still lags behind heuristic methods such as CoreHD in certain scenarios, and its black-box nature limits interpretability.

In contrast, most real-world networks are inherently directed \cite{b1,b2}, including gene regulatory networks \cite{liu2020robustness}, traffic systems \cite{li2015percolation}, and the World Wide Web \cite{strogatz2001exploring}, which often exhibit strong asymmetry and structural heterogeneity \cite{b37}. These features pose challenges for directly applying existing undirected-based dismantling methods.

Fortunately, the recent development of trophic analysis \cite{mackay2020directed} has provided a principled way to quantify directionality and hierarchy in general directed networks, offering interpretable tools to support more effective dismantling strategies.
Trophic analysis was originally developed to study species' trophic levels in ecology \cite{johnson2014trophic,levine1980several} and has since been widely applied across various domains, including food web exploration \cite{klaise2017origin}, epidemic spreading in social networks \cite{klaise2016neurons}, infrastructure resilience assessment \cite{pagani2019resilience, pagani2020quantifying}, and economic interactions \cite{mackay2020directed}. These applications have advanced the understanding of directed networks by providing quantitative tools for analyzing complex systems.

Trophic analysis reveals a strong correlation between a network's strong connectivity, global directionality, and hierarchical organization, quantified by trophic incoherence. Hierarchical structures are prevalent in real-world networks, where nodes’ trophic levels are often indicative of their functional roles \cite{mackay2020directed}. Notably, ``backward" links—where the target node's trophic level is not higher than the source node's—play a crucial role in shaping global directionality. Their removal can significantly disrupt hierarchical ordering and strong connectivity \cite{coupette2022exactly, rodgers2023strong}, presenting a promising avenue for effectively dismantling directed networks.

Based on trophic analysis, we developed an interpretable network dismantling method tailored for directed networks, whose connectivity is characterized by its GSCC size \cite{liu2016breakdown}. In our trophic analysis-based dismantling (TAD) method, nodes are removed in descending order according to their NI centrality scores. When applied to various networks, our TAD method outperforms existing state-of-the-art methods in terms of fragmenting directed networks. Further investigation of the network dismantling process reveals that TAD results in the largest ``avalanche'' compared to other methods, which helps explain its superior performance. In addition, we systematically investigated the general applicability of TAD and find that its performance improves first and then deteriorates as the trophic incoherence rises.  
This work offers a comprehensive analysis and explanation of the observed results, thereby enhancing the understanding of network dismantling across diverse scenarios. 


\begin{figure*}[!ht]
\centering
\includegraphics[width=15cm]{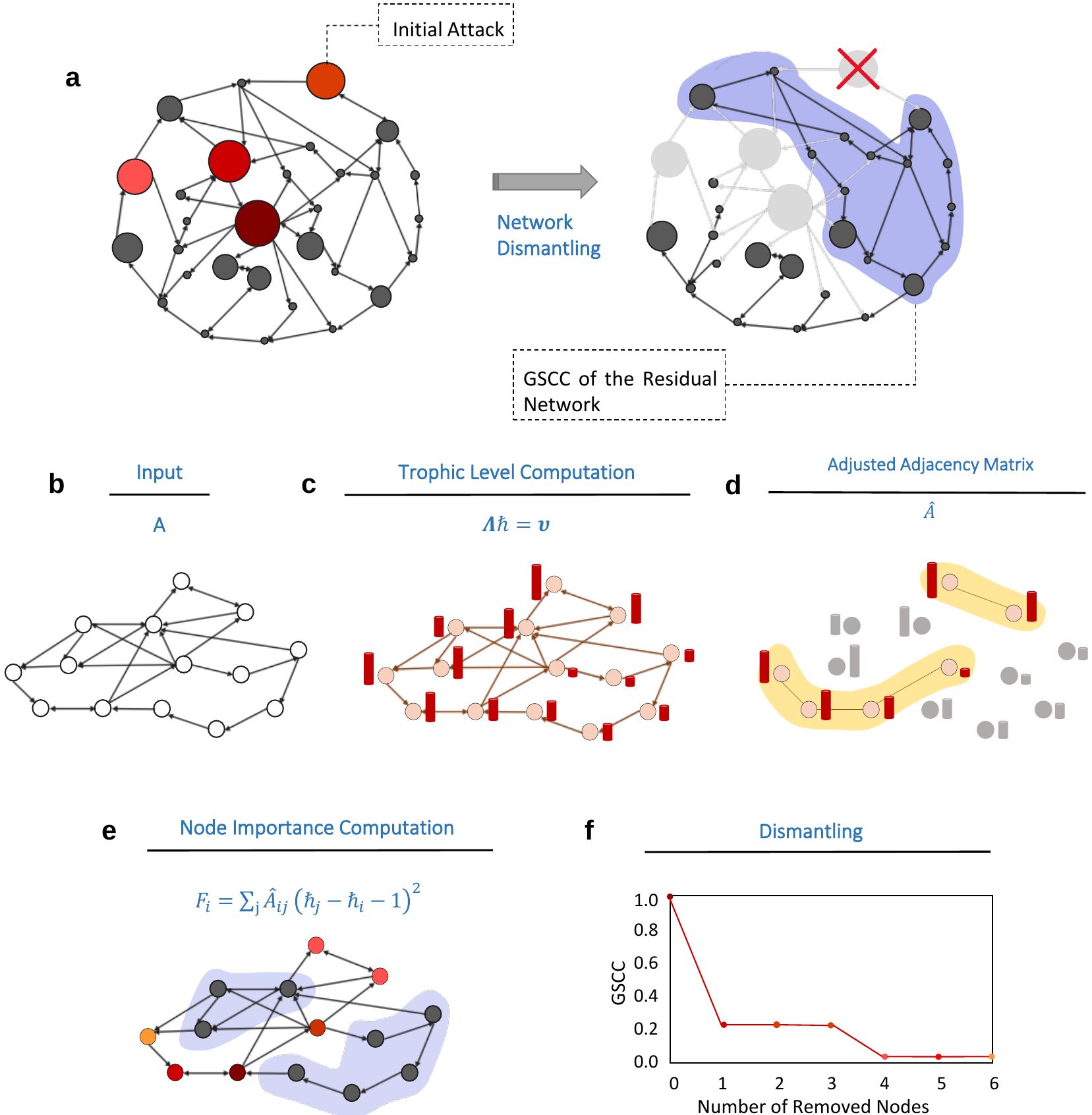}
\vspace{-0.1cm}
\caption{
\textbf{Illustration of the TAD process.}
(\textbf{a}) The network before and after node removal is shown. In the original network, node color (from dark red to light red) represents the attack order, while node size indicates importance. In the residual network, light gray nodes have been removed, and nodes within the purple region constitute the GSCC.
(\textbf{b}–\textbf{e}) Overview of the TAD. The input is the adjacency matrix of the directed network (\textbf{b}). The evaluation of node importance involves three steps: calculating node trophic levels (\textbf{c}); determining the trophic level differences of links and constructing an adjusted undirected network (\textbf{d}); and computing the NI centrality score for each node (\textbf{e}).
(\textbf{f}) Visualization of the dismantling process, where the GSCC size of the residual network decreases as nodes are sequentially removed in the order from dark to light red.
}
\label{stld_fig1_Schema}
\end{figure*}

\begin{figure*}[!ht]
\captionsetup[subfigure]{labelformat=simple}
\centering
\includegraphics[width=16cm]{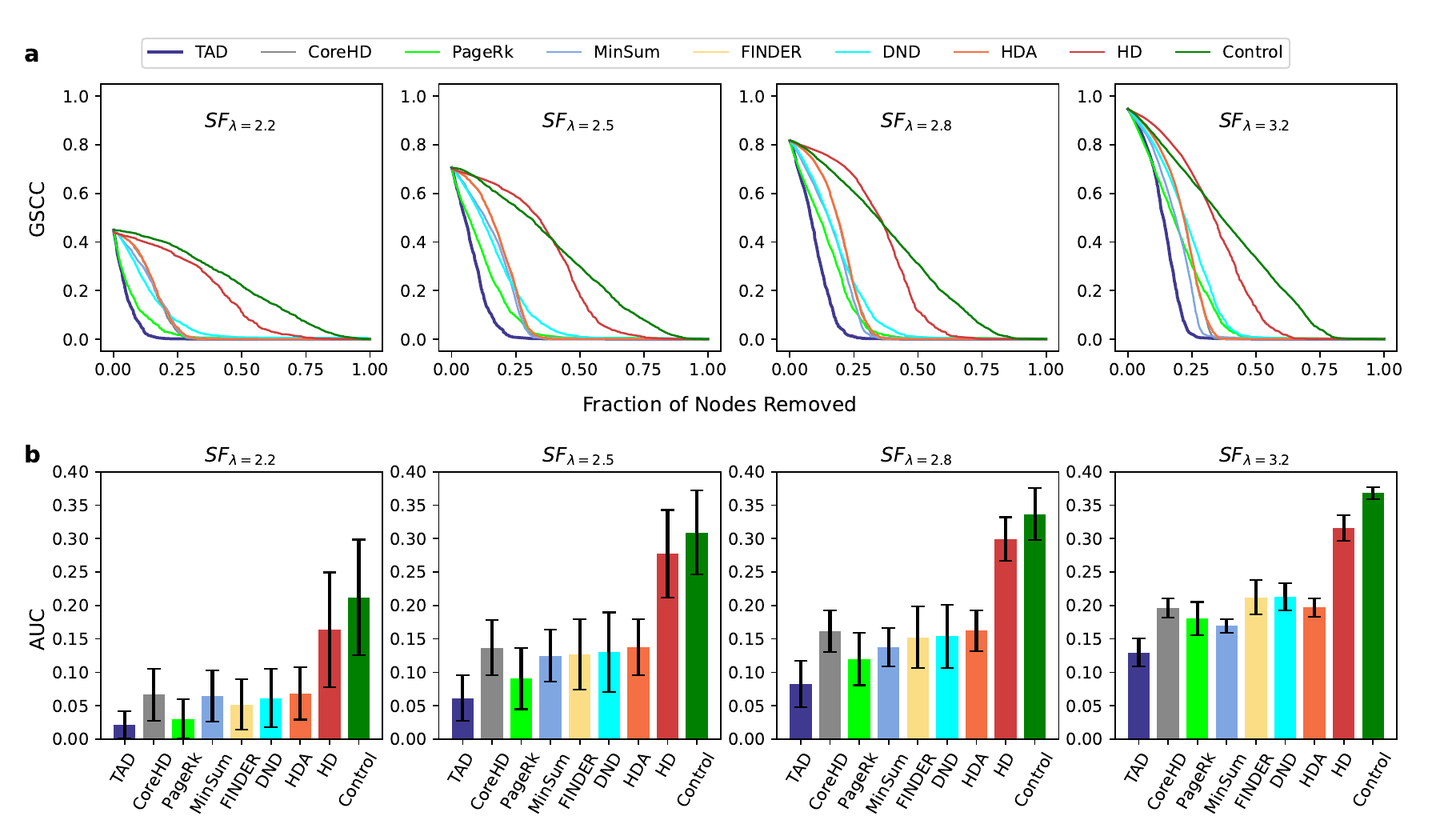}
\caption{
\textbf{Performance of TAD on SF networks with varying degree distribution exponents} ($\lambda$ = 2.2, 2.5, 2.8, 3.2). (\textbf{a}) Comparison of TAD, MinSum, Finder, adaptive degree, and other methods on SF networks. The solid lines indicate the size of the GSCC as a function of the fraction of removed nodes.
(\textbf{b}) The average Area Under the Curve (AUC) of the SF network over the entire dismantling process. The networks contain 1000 nodes and the error bars represent the standard deviations across 30 random realizations.}
\label{stld_fig2_SFBreakdown}
\end{figure*}

\section{Results}

\subsection{Overview of the TAD method}
To address the intricate role of hierarchy and directionality in network robustness, we introduce TAD, a network dismantling method based on trophic level analysis. This method ranks nodes by importance before any removal occurs and then eliminates them in descending order of this importance. The procedure consists of three main steps, as illustrated in Fig.~\ref{stld_fig1_Schema}. Given a directed network, we first compute the trophic level $h_i$ of each node based on the trophic level analysis (see \textit{Methods}).

A directed link $(i \to j)$ is defined as a backward link if the trophic level of its target node is lower than that of its source, i.e., $h_j - h_i < 0$. These backward links are known to play a critical role in maintaining the strong connectivity of the network by disrupting its hierarchical structure \cite{rodgers2023strong}. Therefore, we anticipate that systematically identifying and removing such links is an effective strategy to dismantle the network.

Furthermore, the presence of backward links gives rise to trophic incoherence of the network, which is defined as $F=\frac{\sum_{ij}A_{ij}(h_j-h_i-1)^2}{\sum_{ij}A_{ij}}$, with $A$ being the adjacency matrix \cite{mackay2020directed}. The trophic incoherence of a network $F$ quantifies how much a directed network deviates from a perfectly hierarchical structure. When $F=0$, the network is acyclic, with no backward links. As $F$ increases, the number of backward links grows, introducing structural disorder. Conversely, when $F=1$, all nodes share the same trophic level, as seen in a fully cyclic network. 
 
According to the fact that both trophic incoherence and backward links are key contributors to strong connectivity, we design a dismantling method that prioritizes nodes based on their role in sustaining these structural features. Specifically, we define the NI centrality $F_i$ for each node:
\begin{equation} \label{Fi}
F_i=\sum_j\widehat{A}_{ij}(h_j-h_i-1)^2,
\end{equation}
where $\widehat{A}$ denotes the adjusted adjacency matrix that includes only backward links. In particular, if there is a backward link between nodes $i$ and $j$, then $\widehat{A}_{ij} = \widehat{A}_{ji} = 1$; otherwise, $\widehat{A}_{ij} = 0$. A node with higher NI centrality $F_i$ indicates that it is connected to larger number of weighted backward links, where the term $(h_j-h_i-1)^2$ serves as the weight of these backward links, quantifying each link’s contribution. Alternative weighting functions were also explored (see \textit{Supplementary Information} Sec. S1 for details on the tested variants). This metric effectively captures each node’s influence on the overall incoherence of the network, providing a principled basis for prioritizing node removal during the dismantling process.


\subsection{Dismantling synthetic networks using TAD}

To evaluate the effectiveness of TAD, we compared it against eight baseline methods, including classical centrality-based (HD \cite{b10}, HDA \cite{lu2016vital}, PageRank \cite{gleich2015pagerank}, Control Centrality \cite{liu2012control}), structural optimization (MinSum \cite{minsum}, CoreHD \cite{coreHD}), and machine learning-based approaches (FINDER \cite{finder}, DND \cite{zhang2023dnd}). These methods encompass both classical heuristics and modern AI-driven strategies. Details of each method are presented in the \textit{Methods} section.

Since the functionality of a directed network depends on the integrity of its strongly connected component, we tracked the evolution of the GSCC size during the progressive removal of nodes in synthetic Scale-Free (SF) networks \cite{liu2016breakdown} with varying degree exponents $\lambda$. As shown in Fig.~\ref{stld_fig2_SFBreakdown}A, TAD (dark blue lines) consistently induces a faster collapse of the GSCC compared to other methods, demonstrating its superior dismantling performance.

To quantify dismantling efficiency, we plotted the curve of the remaining GSCC size as a function of the fraction of nodes removed, and then calculated the area under this curve (AUC). A smaller AUC value indicates a more effective dismantling. As presented in Fig.~\ref{stld_fig2_SFBreakdown}B, TAD consistently achieves the lowest AUC across all $\lambda$ values. For instance, when $\lambda = 2.2$, the AUC for TAD is 0.02, significantly better than the next best method, PageRank (0.04). Similarly, at $\lambda = 3.2$, TAD yields an AUC of 0.13, while the second-best result is 0.18.

Moreover, the dismantling advantage of TAD becomes more pronounced as $\lambda$ decreases. A lower $\lambda$ indicates higher heterogeneity in the degree distribution—a typical property of many real-world networks. This trend suggests that TAD is particularly effective in dismantling highly heterogeneous directed networks. 

To further evaluate the generalizability of TAD, we tested its performance on synthetic Erdős–Rényi (ER) networks with varying average degrees. As detailed in \textit{Supplementary Information} Sec.~S2, TAD remains highly effective in ER networks with moderate link density. However, as the average degree increases, the network becomes more cyclic and trophically incoherent, reducing the advantage of hierarchy-based strategies such as TAD—a phenomenon that will be further examined in the next section. These results underscore the structural dependence of TAD’s effectiveness and highlight the importance of considering trophic properties when selecting dismantling strategies.

\subsection{TAD induces the largest maximum avalanche}
In the process of network dismantling, we observed an ``avalanche" phenomenon, which is a first-order-like cascading fragmentation of the network as nodes are removed \cite{Hinrichsen2000}. This phenomenon indicates that the failure of one node can lead to other nodes' failure, causing a significant reduction of the GSCC size. The larger the avalanche, the more effectively the network is dismantled, which suggests that the network's connectivity has been severely damaged, leading to a major breakdown.

\begin{figure*}[!ht]
\captionsetup[subfigure]{labelformat=simple}
\centering
\hspace{-0.5cm}
    \includegraphics[width=17.5cm]{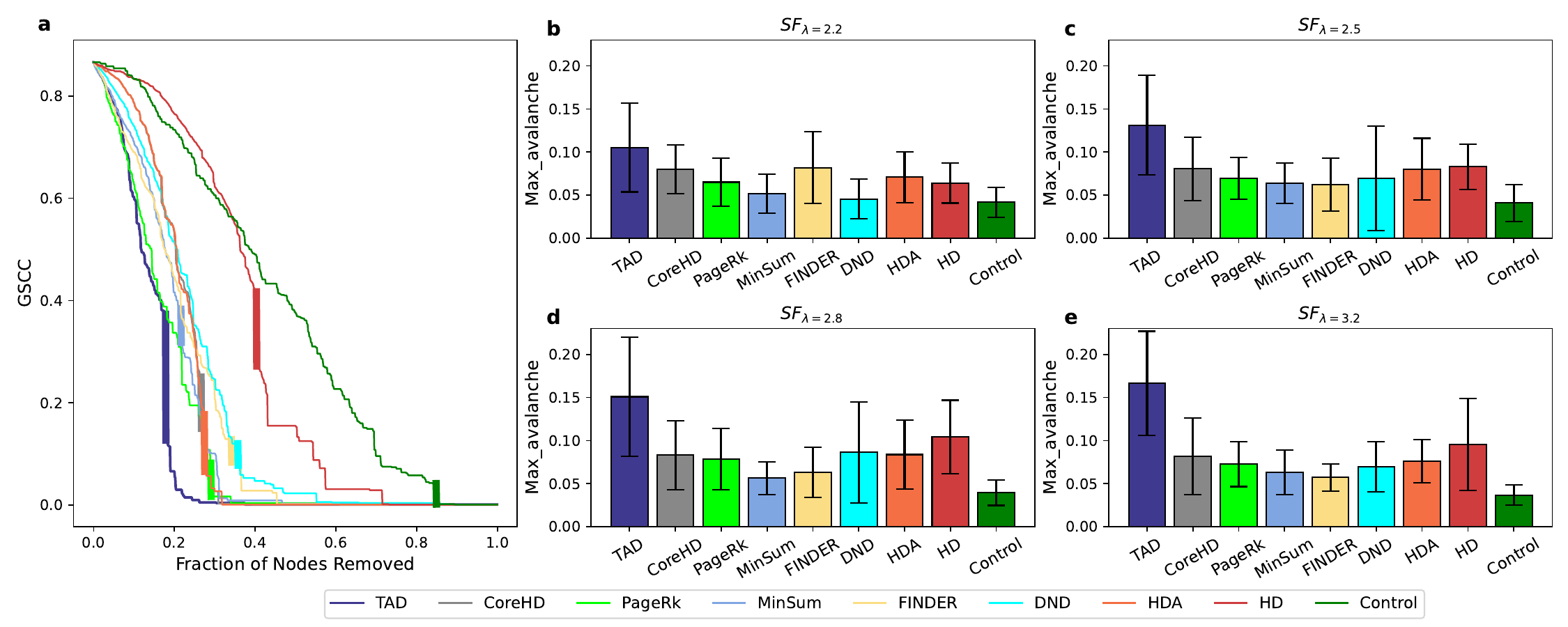}
\vspace{-0.1cm}
\caption{\textbf{The maximum avalanches caused by different methods.} (\textbf{a}) The process of dismantling a directed network using various methods. The horizontal axis represents the fraction of nodes removed, and the vertical axis shows the size of the GSCC. The maximum avalanches are highlighted in thicker lines. (\textbf{b})-(\textbf{e}) The size of the maximum avalanches caused by different methods across SF networks with different degree distribution exponent $\lambda$=2.2, 2.5, 2.8, and 3.2. The results are averaged over 30 realizations.}
\label{stld_fig5_MaxAvalanche}
\end{figure*}

As shown in Fig. \ref{stld_fig5_MaxAvalanche}A, we highlighted the maximum avalanche triggered by different methods. Across different networks, shown as Fig. \ref{stld_fig5_MaxAvalanche}B-E, TAD stands out by consistently inducing the largest maximum avalanche. In contrast to methods such as CoreHD, PageRank, and FINDER, which cause a more gradual and less disruptive fragmentation, TAD leads to a sharper and faster breakdown of the network’s connectivity, demonstrating its superior ability to dismantle directed networks.

TAD induces the largest maximum avalanche due to its strategic node removal, which prioritizes nodes based on their contribution to the overall network incoherence. A backward link connects nodes in a way that opposes the natural hierarchical structure of the network, which is crucial to the formation of GSCC. By targeting nodes with a high number of backward links (significant trophic level differences), TAD effectively dismantles critical connections in the network and leads to abrupt directed percolation. The ability of TAD to induce the largest maximum avalanche is meaningful as it offers a powerful tool for analyzing the vulnerability and resilience of complex networks, shedding light on the potential weaknesses of networks in various real-world systems.

\subsection{Performance of TAD across networks with varying trophic incoherence}

Our TAD method is designed based on node's network incoherence centrality, prioritizing node removal according to the number and trophic level difference of backward links. On the network level, different networks may have different trophic incoherence $F$, which could influence TAD's performance. Investigating how TAD performs across networks with varying $F$ is crucial for assessing its effectiveness and generalizability in different structural configurations.

We compared the performance of TAD with other dismantling methods in networks with varying trophic incoherence $F$, as shown in Figure~\ref{stld_fig6_PerformanceF0}.  
In ER networks with an average degree $\langle k \rangle=20$ (see \textit{Supplementary Information} Sec. S3 for the generation of ER networks with wary $F$), as shown in panel (A), TAD achieves the lowest AUC for networks with $F \leq 0.62$, indicating its superior dismantling capability in networks with moderate hierarchical structure. As $F$ increases beyond this point, the efficiency of TAD gradually decreases. In SF networks with a degree exponent $\lambda=2.8$, shown in panel (B), we analyzed cases where $F \leq 0.6$ and extended the analysis to networks with $F > 0.6$ by reversing some links to artificially increase trophic incoherence (see \textit{Supplementary Information} Sec. S4 for more details). The pattern observed in SF networks is consistent with that in ER networks: TAD exhibits the best performance at low $F$, but its advantage diminishes as $F$ grows. When $F$ exceeds 0.67, TAD is no longer the top-performing method.

This trend can be attributed to the structural differences associated with varying trophic incoherence. In networks with low to moderate $F$, the hierarchical structure is still evident, allowing TAD to effectively exploit trophic-level imbalances for identifying critical nodes and achieving efficient fragmentation. As $F$ increases, the growing prevalence of backward links initially introduces more vulnerabilities. However, in highly incoherent networks, the directional hierarchy becomes obscured, reducing the effectiveness of TAD’s hierarchy-based strategy. These results suggest that TAD is best suited for networks where hierarchy is present but not fully acyclic, and they highlight the importance of structural coherence in guiding dismantling strategies.

\begin{figure*}[!ht]
\captionsetup[subfigure]{labelformat=simple}
\centering
\hspace{-0.5cm}
    \includegraphics[width=15cm]{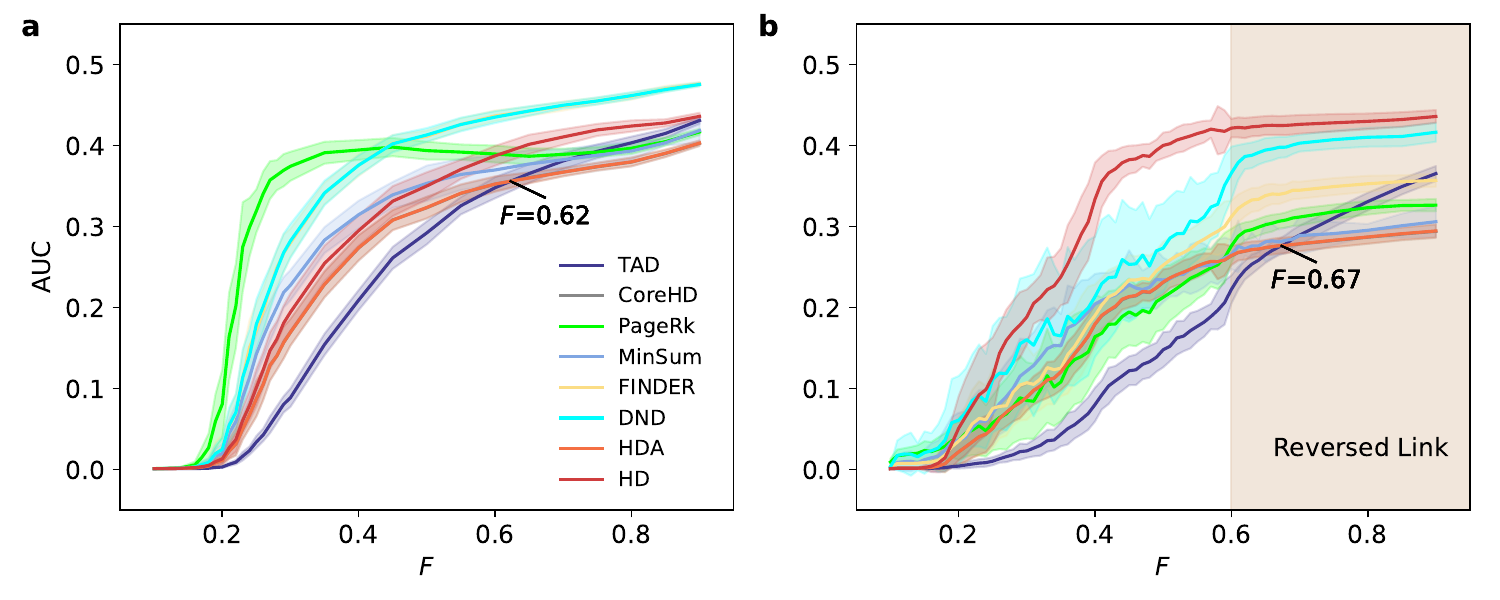}
\vspace{-0.1cm}
\caption{\textbf{The performance of different network dismantling methods across networks with varying $F$}. The horizontal axis represents the trophic incoherence of a network, while the vertical axis corresponds to AUC, where a lower AUC indicates greater dismantling efficiency. 
(\textbf{a}) The results for ER networks ($\langle k \rangle=20$) generated with varying trophic incoherence using a simulated annealing method \cite{rodgers2022network}. The results show that TAD achieves the lowest AUC when $F$ is below 0.62, indicating that it is the most effective dismantling method in networks with moderate hierarchical structure.
(\textbf{b}) Results for SF networks ($\lambda = 2.8$) across different levels of trophic incoherence $F$. Since randomly generated SF networks rarely reach $F > 0.6$, we artificially increased $F$ by reversing a fraction of links. TAD exhibits the best performance (lowest AUC) when $F$ is small, confirming its effectiveness in networks with strong hierarchical structure. As $F$ increases beyond $0.67$, its AUC continues to rise and is eventually surpassed by other methods. This transition highlights the structural dependence of dismantling strategies and suggests that TAD is particularly suited to networks with moderate to low incoherence, where hierarchy plays a central role.
}
\label{stld_fig6_PerformanceF0}
\end{figure*}

\subsection{Applying TAD to real-world directed networks}

To evaluate the practical utility of TAD, we applied it to 15 real-world directed networks spanning diverse domains, including food webs, neural networks, scholarly citations, social networks, infrastructure networks, and global trade systems. The characteristics of real-world directed networks are shown in Table \ref{realnetworks}, where $N$ and $L$ respectively represent the number of nodes and links, and $GSCC_0$ is the initial GSCC size. These networks differ significantly in size and structural properties, offering a comprehensive test of method generalizability.

\begin{figure*}[!ht]
\captionsetup[subfigure]{labelformat=simple}
\centering
\hspace{-0.5cm}
    \includegraphics[width=16cm]{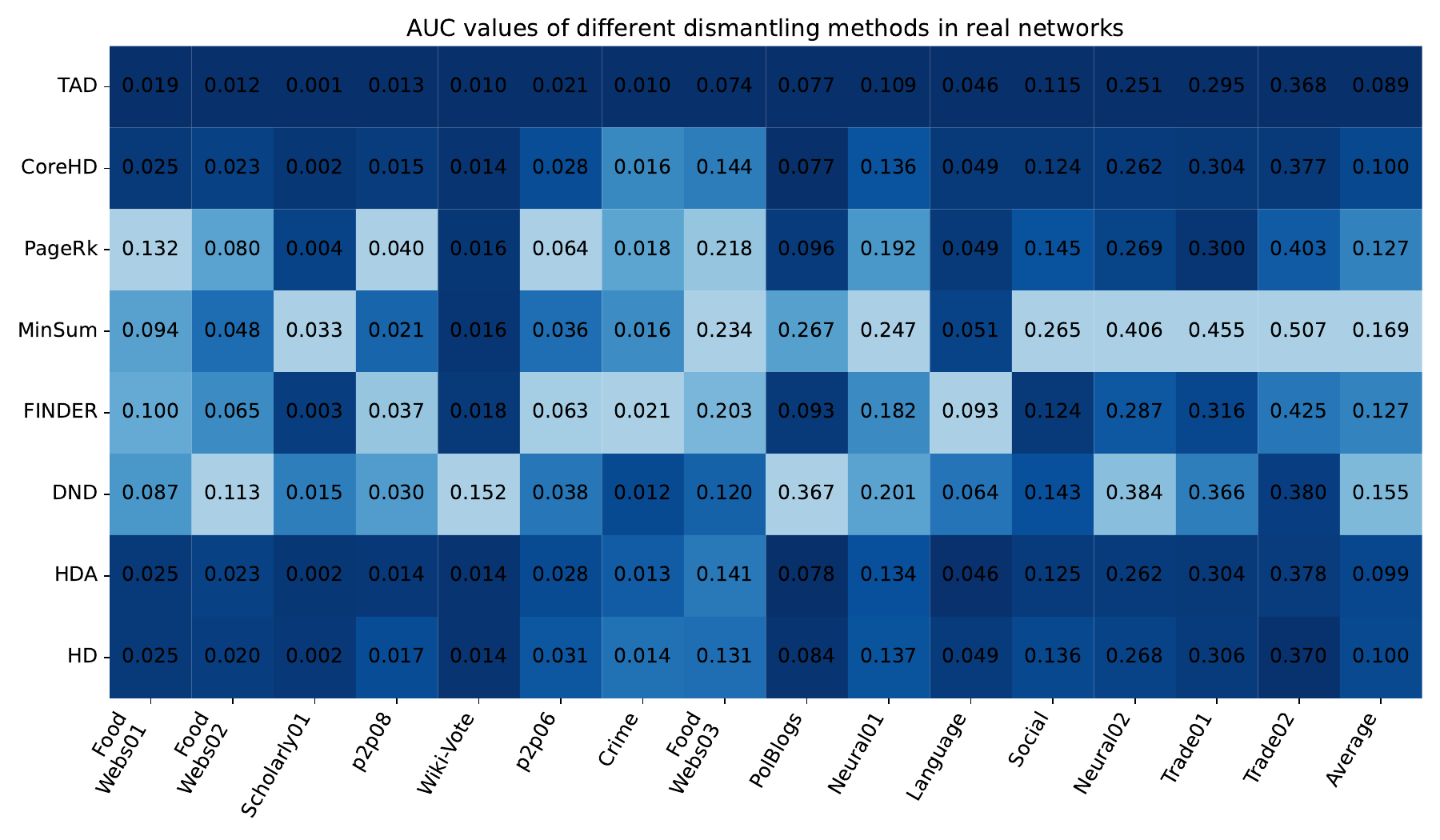}
\vspace{-0.1cm}
\caption{
\textbf{AUC values of different dismantling methods applied to 15 real-world directed networks.} The horizontal axis lists the networks from various domains (e.g., ecological, neural, scholarly, and social), and the vertical axis shows the average area under the GSCC curve (AUC) for each method. Lower AUC values indicate better dismantling performance. TAD consistently achieves the lowest AUC in all networks,
demonstrating its superior effectiveness in fragmenting real-world directed networks. The final column shows the average AUC across all networks. }
\label{stld_fig3_RealBreakdown}
\end{figure*}

Figure~\ref{stld_fig3_RealBreakdown} presents the AUC values of the dismantling curves for all methods across 15 real-world networks. TAD consistently achieves the lowest AUC, demonstrating its superior performance compared to baseline methods (see \textit{Supplementary Information} Sec. S5 for more details). Notably, unlike in synthetic networks, TAD remains the most effective even in real networks with very high trophic incoherence $F$.

To gain further insight into the dismantling dynamics, we visualized the dismantling process of two representative networks: \texttt{NeuralNet\_C.elegans} and \texttt{FoodWebs\_reef} (Fig.~\ref{stld_fig4_Real Example}). In both cases, TAD rapidly breaks down the GSCC with significantly fewer node removals than competing methods. For example, in \texttt{NeuralNet\_C.elegans}, the GSCC is reduced to 21.5\% after removing just 18\% of the nodes. Similarly, in \texttt{FoodWebs\_reef}, the GSCC drops to 8.0\% after 18\% of nodes are removed. Snapshots at different dismantling stages (panels B–D and F–H) show that TAD effectively targets structurally critical nodes, leading to abrupt transitions in connectivity.

The superior performance of TAD stems from its ability to exploit hierarchical directionality in real-world networks. High-scoring nodes in the NI centrality ranking often exhibit distinct topological patterns: they are typically located at junctions between high and low trophic levels, either concentrating or redistributing structural flow. These nodes may be (i) low-level nodes controlled by many higher-level nodes, (ii) high-level nodes connecting downward, (iii) nodes bridging hierarchical gaps, or (iv) hubs in backward-link chains. The removal of such nodes triggers major disruptions in the GSCC structure.

These findings demonstrate that TAD not only excels quantitatively across real-world networks, but also captures mechanistic insights into which nodes sustain directed connectivity. The method leverages directional hierarchy as a structural prior, making it especially effective in real-world systems where directionality encodes functional constraints.

\begin{table}[ht!]
\caption{
Characteristics of real-world directed networks.
}
\centering
\begin{tabular}{p{3.2cm}p{1.2cm}p{1.2cm}p{1.5cm}p{1.3cm}p{6cm}}
\hline
Network & $N$ & $L$ & $GSCC_0$ & $F$    & Description \\
\hline

Food Webs01 \cite{johnson2020data}      & 92    & 997    & 0.228 & 0.208 & Little Rock food web        \\
Food Webs02 \cite{johnson2020data}    & 483   & 15362  & 0.203 & 0.217 & Weddel Sea food web         \\
Scholarly01 \cite{mccallum2000automating}     & 23166 & 91500  & 0.172 & 0.234 & Citation network (Cora)\\
P2p08 \cite{ripeanu2002mapping}         &  6301  & 20777  & 0.328 & 0.251 & Gnutella P2P network (Aug 8, 2002)\\
Wiki-Vote \cite{leskovec2010predicting}  & 7115  & 103689 & 0.183 & 0.283    & Wikipedia vote network     \\
P2p06 \cite{ripeanu2002mapping}          & 8717  & 31525  & 0.37  & 0.297 & Gnutella P2P network (Aug 6, 2002) \\
Crime \cite{crime}  & 829   & 1476   & 0.505 & 0.337 & Moreno's crime network            \\
Food Webs03 \cite{johnson2020data}    & 50    & 556    & 0.56  & 0.373  & Reef food web               \\
PolBlogs \cite{adamic2005political}    & 1224   & 19022    & 0.648  & 0.448  & Political blogs               \\
Neural01 \cite{johnson2020data} & 297   & 2345   & 0.805 & 0.498  & Neural network of C. Elegans       \\
Language \cite{johnson2020data} & 50    & 101    & 0.52  & 0.573 & Word sequence network         \\
Social \cite{johnson2020data}         & 32    & 96     & 0.719 & 0.601  & Student social network  \\
Neural02 \cite{johnson2020data}     & 242   & 4090   & 1     & 0.793  & Neural network of Rhesus monkey            \\
Trade01 \cite{johnson2020data}  & 24    & 310    & 0.917 & 0.842 & Trade network of basic goods              \\
Trade02 \cite{johnson2020data} & 24    & 307    & 1     & 0.855  & Trade network of food                \\

\hline
\end{tabular}
\label{realnetworks}
\end{table}

\begin{figure*}[!ht]
\captionsetup[subfigure]{labelformat=simple}
\centering
\hspace{-0.5cm}
    \includegraphics[width=17.5cm]{fig/stld_fig4_Real_Example.pdf}
\vspace{-0.1cm}
\caption{\textbf{Dismantling processes of two real-world directed networks using TAD and baseline methods.} 
Panels (\textbf{a}–\textbf{d}) illustrate the dismantling process of the \texttt{NeuralNet\_C.elegans} network.
(\textbf{a}) The GSCC size declines as more nodes are removed. TAD (blue curve) consistently outperforms baseline methods by more effectively reducing the GSCC.
(\textbf{b}–\textbf{d}) Network snapshots corresponding to 5\%, 14\%, and 18\% node removal, respectively. Red and pink nodes belong to the current GSCC, where red denotes the next node to be removed, and pink indicates the remaining GSCC members. The blue nodes represent those that have been separated from the GSCC. Node size reflects the NI centrality scores.
Panels (\textbf{e}–\textbf{h}) present analogous results for the \texttt{FoodWebs\_reef} network.
(\textbf{e}) The GSCC size curve again highlights the superior dismantling performance of TAD.
(\textbf{f}–\textbf{h}) Snapshots at 5\%, 12\%, and 18\% node removal show the GSCC shrinking from 46.0\% to 14.0\% and finally to 8.0\%.}
\label{stld_fig4_Real Example}
\end{figure*}

\section{Discussion}

Dismantling directed networks is a fundamental problem with wide-ranging applications, from epidemic control and rumor suppression to infrastructure protection and social system regulation. Unlike undirected networks, directed networks are governed by asymmetric interactions and hierarchical organization, making their structural fragility and functional resilience more complex to assess. Existing dismantling methods, largely developed for undirected graphs, often fall short in capturing these features.

To address this gap, we developed the TAD method—an interpretable and efficient dismantling strategy tailored for directed networks. Grounded in trophic level analysis, TAD prioritizes node removal based on the network incoherence centrality, a measure to quantify the trophic level difference across backward links, thereby targeting the nodes that most disrupt the network's hierarchical integrity. This mechanism leverages directional and hierarchical cues that are often overlooked by traditional or learning-based methods.

Our results demonstrate that TAD consistently outperforms state-of-the-art dismantling methods across both synthetic and real-world networks, particularly in systems with high degree heterogeneity or well-defined global directionality. In real-world directed networks from multiple domains, TAD achieves the lowest dismantling AUC in all cases. Notably, it induces the largest maximum avalanches, revealing its capacity to identify structurally critical nodes whose removal causes cascading breakdowns in connectivity.

A key insight from this work is the structural dependence of dismantling effectiveness. In synthetic networks, TAD performs best when trophic incoherence is low to moderate—where hierarchical organization remains but is not entirely acyclic. As incoherence increases and the network becomes more cyclic and disordered, the relative advantage of hierarchy-based strategies gradually declines, emphasizing the need to tailor dismantling approaches to different structural regimes. Remarkably, in contrast to synthetic cases, TAD maintains its superiority in real-world networks, consistently outperforming all baseline methods even when the trophic incoherence $F$ is very high.

Beyond its empirical performance, TAD offers mechanistic interpretability by identifying nodes that bridge trophic levels, bottleneck directional flow, or anchor backward connectivity. The avalanche-like collapses it triggers resemble first-order transitions in non-equilibrium systems and bear similarity to directed percolation processes \cite{domany1984equivalence}. This connection provides a physical perspective on the dismantling dynamics and enriches our understanding of robustness in directed networks.

In summary, TAD not only advances the dismantling of directed networks by integrating hierarchical structure into the dismantling strategy, but also provides a generalizable framework to explore the structure-function relationship in complex systems. Its ability to predict and induce critical transitions further enhances its applicability to domains that require precise, efficient, and interpretable control over networked systems.







\section{Methods}


\subsection{Network Dismantling}

Given a directed network $\mathcal{G} = (\mathcal{V}, \mathcal{E})$ with $N = |\mathcal{V}|$ nodes and $|\mathcal{E}|$ directed links, the goal of network dismantling is to identify a set of nodes $\mathcal{S} \subseteq \mathcal{V}$ whose removal completely fragments the network, such that the resulting structure contains no giant strongly connected component (GSCC). 
To assess the effectiveness of a dismantling strategy, we evaluate how rapidly the GSCC collapses as nodes are sequentially removed. Specifically, after removing $k$ nodes $\{v_1, v_2, ..., v_k\}$, we define the remaining connectivity as $f_k = |GSCC(\mathcal{G} \setminus \{v_1, ..., v_k\})| / N$. The overall dismantling performance is measured by the AUC:

\begin{equation}
\label{eq:AUC}
\mathrm{AUC}(\mathcal{G}) = \frac{1}{N} \sum_{k=1}^{K} f_k,
\end{equation}

where $K$ is the total number of node removals required to eliminate the GSCC. A smaller AUC indicates a more efficient dismantling strategy that disrupts strong connectivity with fewer removals. 

As illustrated in Fig.~\ref{stld_fig1_Schema}, the dismantling procedure proceeds by ranking nodes using a given method and removing them sequentially, monitoring the GSCC size after each step. In this work, we use AUC as the unified metric to compare the performance of TAD and other baseline methods across a variety of directed networks.

\subsection{Computation of trophic level}
Trophic levels define the hierarchical position of individual nodes within a directed network. For example, in a food web \cite{johnson2014trophic}, energy flows from nodes at lower trophic levels, such as plants, to those at higher levels, like carnivores. This concept can be applied to various directed networks. In a network with $N$ nodes, $A$ represents the adjacency matrix. If there is a link from node $i$ to node $j$, then $A_{ij}=1$. Otherwise, $A_{ij}=0$. The nodes' trophic levels are determined by solving the matrix equation:
\begin{equation} \label{trophic}
\Lambda h=\upsilon ,
\end{equation}
where $h$ is the vector of trophic levels, $\upsilon_i=k_i^{\rm in}-k_i^{\rm out}$ is the difference between in- and out-degrees, and $\Lambda={\rm diag}(u)-A-A^T$ is the Laplacian matrix of the network $\mathcal G$. Here, $u_i=k_i^{\rm in}+k_i^{\rm out}$ is the sum of in- and out-degree. These quantities can be easily computed by matrix operations. Since the matrix $\Lambda$ is singular, we cannot directly invert it to obtain $h$. However, methods such as LU decomposition or iterative solvers can be used for large networks. For convenience, we follow the convention that the lowest level node is assigned a value of $h = 0$.

 \subsection{Generate networks with high level of trophic incoherence}

In synthetic SF networks generated with a fixed degree distribution exponent $\lambda$, the trophic incoherence $F$ rarely exceeds 0.6 due to the natural emergence of globally directed structures. However, to systematically evaluate the performance of dismantling methods in networks with higher structural disorder, we construct SF networks with elevated trophic incoherence while preserving their degree distributions.

To achieve this, we adopt a minimal perturbation strategy that reverses the direction of a small number of selected links to increase $F$. Specifically, we iteratively select one directed link $(i \rightarrow j)$ and reverse its direction $(j \rightarrow i)$ only if the operation increases the number of cycles in the network, thereby increasing the global incoherence. This process allows us to gradually increase $F$ without significantly altering other topological features, such as degree distribution or density.

\subsection{Baseline methods}

To benchmark the performance of TAD, we compare it against eight representative dismantling methods:
\begin{itemize}
  \item \textbf{HD (High Degree)} \cite{b10}: remove nodes in descending order of their degree in the original network.
  \item \textbf{HDA (High Degree Adaptive)} \cite{lu2016vital}: iteratively remove the node with the highest current degree, updating degrees after each removal.
  \item \textbf{PageRank} \cite{gleich2015pagerank}: rank nodes according to their PageRank centrality and removes them in descending order.
  \item \textbf{MinSum} \cite{minsum}: apply a message-passing strategy to iteratively identify nodes that minimize the size of the largest connected component.
  \item \textbf{CoreHD} \cite{coreHD}: remove the highest-degree node in the network’s 2-core until the core collapses, followed by dismantling the residual tree-like structure. In directed networks, the 2-core is defined based on total degree.
  \item \textbf{Control Centrality} \cite{liu2012control}: prioritize nodes based on their contribution to the controllability of the system.
  \item \textbf{FINDER} \cite{finder}: use a deep reinforcement learning framework to select key nodes for removal. The model is adapted here for directed networks to reduce the size of the GSCC.
  \item \textbf{DND} \cite{zhang2023dnd}: a deep learning-based method designed for directed network disintegration using graph attention mechanisms.
\end{itemize}

\section{Acknowledgements}
This work was supported by the National Natural Science Foundation of China under Grants 62225306, T2422010, 62172170, and U2141235, and by the ``Fundamental Research Funds for the Central Universities”.

\bibliography{Lethality}

\end{document}